\documentclass[11pt]{iopart}

\usepackage{iopams}  
\usepackage{epsfig,amsfonts,amsbsy}
\newcommand{\bra}{\left \langle}
\newcommand{\ket}{\right \rangle}
\newcommand{\pder}[2]{\frac{\partial #1}{\partial  #2}}

\newcommand{\ep}{\epsilon}

\begin{document}
\title{Possible extended forms of thermodynamic entropy}
\author{Shin-ichi Sasa}
\ead{sasa@scphys.kyoto-u.ac.jp}
\address{
Department of Physics, Kyoto University, Kyoto 606-8502, Japan}
\date{\today}

\begin{abstract}
Thermodynamic entropy is determined by a heat measurement  
through the Clausius equality. The entropy then formalizes
a fundamental limitation of operations by the second law
of thermodynamics. The entropy is also expressed as the 
Shannon entropy of the microscopic degrees of freedom. 
Whenever an extension of thermodynamic entropy is attempted,
we must pay special attention to how its  three different 
aspects just mentioned are altered. In this paper, we discuss  
possible extensions of the thermodynamic entropy.
\end{abstract}




\section{Introduction}



Equilibrium entropy was introduced as a state variable associated 
with quasi-static heat \cite{Fermi}.
We here review a construction method by Caratheodory \cite{Caratheodory},
where quasi-static heat $d'Q$ is first defined along a path element 
in a state space.  For example, for a simple fluid characterized by
the internal energy $U$ and the volume $V$, we start with 
$d'Q \equiv  dU-pdV$ in the state space consisting of equilibrium states
represented by $A = (U,V)$, where $p(A)$ is the pressure.
Then, under the assumption that $d'Q  = 0$ defines  
curves in  the two-dimensional state space 
(or  generally $d-1$-dimensional surfaces 
in a  $d$-dimensional state space), it turns out that there exists 
a state variable $S$ with an  integration factor $T$ that satisfies 
$dS=d'Q/T$, where $S$ and $T$ are determined in an essentially unique manner.  
This means that, for two different equilibrium states $A$ and $B$, 
\begin{equation}
S(B)-S(A) =  \int_A^B \frac{d' Q}{T}
\label{Clausius-eq}
\end{equation}
holds for any paths from $A$ to $B$ in the state space. 
The state variable $S$ determined by this procedure is 
the thermodynamic entropy, and $T$ the temperature.
It should be noted that paths in the state space,
which are referred to as quasi-static processes, can be 
realized by controlling thermodynamic parameters quite slowly.
With the introduction of 
the entropy, the fundamental relation in thermodynamics 
\begin{equation}
dU=TdS-pdV
\label{fundamental}
\end{equation}
is obtained. The relation describes  material properties
in a unified manner so that we can easily understand a 
non-trivial fact  that the volume dependence of a
heat capacity with $T$ fixed is determined from the 
temperature  dependence of a pressure with the volume fixed. 


For more general time dependent processes  from an
equilibrium state $A$ to another equilibrium state $B$,
the inequality 
\begin{equation}
S(B)-S(A)  \ge  \int dt  \frac{J(t)}{T(t)}
\label{Clausius-ineq}
\end{equation}
holds, where $T(t)$ is the temperature of a single heat 
bath in contact with a system, and $J(t)$ is a heat flux
(i.e. the energy transfer from the heat bath) at time $t$.  
The inequality describes a fundamental 
limitation of operations, which is one 
expression of the second law  of thermodynamics. 
Furthermore, let $P_{\rm eq}(\Gamma)$ be a probability density
of microscopic state $\Gamma$ in an  equilibrium state $A$. 
Then, according to statistical mechanics, $S(A)$ 
is equal to the Shannon entropy of $P_{\rm eq}(\Gamma)$ defined by
\begin{equation}
S(A)\equiv  - \int d\Gamma P_{\rm eq}(\Gamma) \log P_{\rm eq}(\Gamma) .
\label{Shannon}
\end{equation}


The entropy in the elegant formulation described above
is defined for equilibrium states. 
It is quite natural to question the  possibility of entropy extended to 
non-equilibrium states. Indeed, several attempts were proposed 
in the twentieth century
\cite{JouCasasVazquezLebon01,Keizer87,Eu98,EvansHanley80b,Landauer,Oono}.
Here, although we  can define our entropy as we like, the heart of 
the problem is to study whether or not the entropy by some definition 
exhibits other (apparently unrelated) aspects, as for the standard 
thermodynamic entropy. For example, if an extended entropy is defined 
from statistical properties, it is not obvious at all to have a relation 
with heat, which may be interpreted as an extended Clausius relation.


In any case, if we attempt to extend the thermodynamic entropy, we 
first have to consider an extension of the state space so that it 
involves non-equilibrium steady states in addition to equilibrium states. 
As the simplest case, we assume that a driving force is applied to a 
system in contact with a single heat bath of temperature $T$ fixed. 
In this case, the non-equilibrium axis $X$ (e.g. driving force,
 shear rate,  or current) is added to the state space 
of equilibrium states.  We then assume that ${\cal A}=(A, X)$
represents a steady state and we consider a collection of ${\cal A}$
as the extended state space.  That is, time-dependent states are
excluded from the state space. 
Our question is now expressed as follows.
{\it Find a state variable ${\cal S}({\cal A})$ satisfying 
a natural extension of the relations (\ref{Clausius-eq}), 
(\ref{fundamental}), (\ref{Clausius-ineq}), and (\ref{Shannon}).}  


Along with a naive expectation that the Shannon entropy still plays
an important role in non-equilibrium steady states, we assume that 
${\cal S}$ is given by the Shannon entropy,
\begin{equation}
S({\cal A})= - \int d\Gamma P_{\rm st}(\Gamma) \log P_{\rm st}(\Gamma) .
\label{Shannon-ss}
\end{equation}
Then, it can be shown that 
\begin{equation}
S({\cal B})-S({\cal A})  \ge  \int dt \frac{J(t)}{T}.
\label{Clausius-ineq-ss}
\end{equation}
Indeed, such inequality has been understood well since the 
discovery of the fluctuation theorem \cite{Evans, Gallavotti,
Kurchan, Maes, FT-exp, Seifert} and the Jarzynski equality 
\cite{Jarzynski, Crooks,J-exp,C-exp}. However, the equality 
in (\ref{Clausius-ineq-ss}) does not hold even for processes
realized by quasi-static operations, 
simply because the entropy production is positive
in non-equilibrium steady states. In other words,  
$d'Q$ is not defined (divergent). Therefore, 
the Shannon entropy in non-equilibrium steady states is not 
directly related to the heat.


In order to have an extended Clausius equality for quasi-static
operations, we consider a modified heat or a renormalized heat 
$Q^{\rm ren}$ such that 
$d'Q^{\rm ren}$ is well-defined along a path in the extended state space.
Among several possibilities, the simplest choice of $Q^{\rm ren}$
may be the excess heat  $Q^{\rm ex}$ defined by
\begin{equation}
Q^{\rm ex}=\int dt [J(t)-J_{\rm st}(\alpha(t))],
\label{excess}
\end{equation}
where  $J_{\rm st}(\alpha)$ is the expectation value of the 
heat flux in the steady state of the system with the fixed 
parameter $\alpha$ (such as volume, temperature, and external force). 
By changing the parameter quite slowly, we expect that $d'Q^{\rm ex}$
can be defined. Indeed, by considering an infinitely small step operation
for specific mathematical models, we can confirm that $d'Q^{\rm ex}$ 
is well-defined. Then, if this $d'Q^{\rm ex}$ is a proper extension of $d'Q$,
it could be expected that there exists a state variable ${\cal S}$
satisfying 
\begin{equation}
{\cal S}({\cal B})-{\cal S}({\cal A}) 
=  \int_{\cal A}^{\cal B} \frac{d' Q^{\rm ex}}{T}
\label{Clausius-oono}
\end{equation} 
for any quasi-static operations. 



The phenomenological framework of thermodynamics on the basis of 
$d'Q^{\rm ex}$ was investigated by Oono and Paniconi \cite{Oono}.
The validity of this proposal can be confirmed by calculating the 
value of $\int d'Q^{\rm ex}/T $ along closed paths in the extended 
state space. If the integration value always zero, which corresponds
to the integrability condition, $d'Q^{\rm ex}/T$  
can define a  state variable by its integration along a path. 
This mathematical proposition was addressed by  Ruelle  \cite{Ruelle} 
who studied an isokinetic model, independently of Oono and Paniconi.
He  pointed out that if the system is in the linear response regime,
the integrability condition is satisfied and that the state variable
defined through the condition is equivalent to (\ref{Shannon-ss}).
After that, for a very wide class of systems, but still near 
equilibrium, the state variable can be constructed by this 
type of  extended Clausius equality \cite{KNST,KNST2,Nakagawa}, 
while the statistical expression of the entropy is modified 
as the symmetrized Shannon entropy for general cases. 
(See (\ref{Shannon-ss-2}) for its expression.)
 The integrability condition becomes more 
evident in the geometrical formulation 
for Markov jump processes \cite{Sagawa,Yuge}.  


Formally speaking, there is no particular reason for the claim that 
$d'Q^{\rm ex }$ is a proper extension of $d'Q$. Recently, new 
types of extended Clausius inequality have been proposed
\cite{Christian, Jona-Lasinio} without employing the excess heat (\ref{excess}). 
With regard to such modification, it should be recalled that Hatano and Sasa 
introduced a renormalized heat, which was referred to as ``excess heat'', 
but takes a different form from (\ref{excess}), and that they formulated
a generalized second law on the basis of the renormalized heat
\cite{Hatano-Sasa}. In this paper, we discuss possible 
extensions of thermodynamic entropy from a viewpoint of the Hatano-Sasa
relation. 


This paper is organized as follows. In section \ref{model},
for simplicity, we introduce a simple Langevin model that 
describes non-equilibrium Brownian motion. Note however that
our arguments can apply to general Markov  processes. After 
quickly reviewing the second law of equilibrium thermodynamics,
we address the question explicitly in the model.  In section 
\ref{result}, we derive the Hatano-Sasa relation with emphasis
of a role of the dual system. A non-trivial nature of systems with
odd parity variables is also understood from the argument, and
the extended Clausius equality with the excess heat is re-derived 
in a straightforward manner.


\section{Question} \label{model}



If there exists an extension of the thermodynamic framework,
it should apply to a wide class of non-equilibrium systems
such as heat conduction systems, sheared systems, molecular 
motors, and so on. 
The description depends on  systems we  study. For example, 
heat conduction systems are described by Hamiltonian equations
supplemented with stochastic reservoirs, and molecular motors
are described by Langevin equations with Markov processes for
chemical reaction. Thus, an extension of the thermodynamic 
framework should be considered  at least for these systems 
in a universal manner.  Keeping such a universal feature in mind, 
however, we focus on the simplest example: a Langevin equation
\begin{equation}
\gamma \dot x= f-\pder{U(x;\nu)}{x}+\xi
\label{Lan}
\end{equation}
that describes a non-equilibrium Brownian motion on a circuit of length
$L$, where the dot symbol on the top of $x$  
represents the time derivative, $\gamma$ a friction constant,  
$f$ a uniform driving force, $U(x;\nu)$ a $L$-periodic potential 
with a parameter $\nu$ that characterizes the shape of the potential,
and $\xi$  noise   that satisfies
\begin{equation}
\bra \xi(t)\xi(t') \ket =2 \gamma T \delta(t-t').
\end{equation}
We set $\alpha =(f,\nu)$ and change the value of $\alpha$ in time.
We express the protocol as a  function of time 
$\hat\alpha \equiv (\alpha(t))_{t=-\tau}^\tau$ with setting 
$\alpha(-\tau)=\alpha_{\rm i}$ and $\alpha(\tau)=\alpha_{\rm f}$. 
Here, we assume that the external operation is done only in the 
interval $[-\tau', \tau']$ with $0 \le \tau' \ll \tau$. 
More precisely,  $\tau-\tau'$ is chosen to be much longer 
than a relaxation time of the system. 


Although we mainly analyze (\ref{Lan}) in this paper, the argument 
is not restricted to the specific model. Indeed, it is easy to 
replace the argument so as to investigate general Markov processes.
In order to see the correspondence with general cases, we 
introduce  a  discrete time as $t_k =-\tau+ 2k\tau/K$,
$ 0 \le k \le K $. We assume the protocol as
$\alpha(t)= \alpha_k $  for $t_{k} \le t \le t_{k+1}$
with  $0 \le k \le K -1$.  Let $\Psi(x_k \to x_{k+1};\alpha_k)$
be a transition probability to $x_{k+1}$ from $x_k$ in a time
interval $[t_k, t_{k+1}]$. 
(See (\ref{Psi}) for the Langevin equation (\ref{Lan}).)
Suppose that the heat 
$Q(x_k \to x_{k+1};\alpha_k) $ is determined from 
an energetic consideration. (See  (\ref{heat-ness}) 
for the Langevin equation (\ref{Lan}).) 
The results presented below are derived for models that satisfy 
\begin{equation}
-\beta Q(x_k \to x_{k+1};\alpha_k) 
= \log \frac{\Psi(x_k \to x_{k+1};\alpha_k)}
{\Psi(x_{k+1} \to x_{k};\alpha_k)},
\label{LDB}
\end{equation}
which is called the {\it local detailed balance condition} \cite{d-FT}. 
It can be confirmed that the Langevin equation with the energetic 
interpretation satisfies the local detailed balance condition. 
For other interesting non-equilibrium steady states for systems
in contact with multiple heat baths, (\ref{LDB}) should be replaced
by a different form. Corresponding to this replacement, the argument
below also should be modified. 


We first review the second law of thermodynamics for
the model (\ref{Lan}) with $f=0$. The stationary probability 
density of $x$, $P_{\rm st}(x;\nu)$, is expressed as
\begin{equation}
P^{\rm can}_\beta (x;\nu)=\exp[-\beta (U(x;\nu)-F(\beta,\nu)],
\label{can}
\end{equation}
where $F$ corresponds to the free energy determined 
from the normalization of probability and $\beta=1/T$. 
The Boltzmann constant is set to unity.
The Shannon entropy $S$ in the equilibrium state is 
defined as
\begin{equation}
S(\nu) \equiv - \int dx P^{\rm can}_\beta(x;\nu) 
\log P^{\rm can}_\beta(x;\nu).
\end{equation}
Since $f=0$, $\hat \nu=(\nu(t))_{t=-\tau}^\tau$
with $\nu(-\tau)=\nu_{\rm i}$ and $\nu(\tau)=\nu_{\rm f}$ is used
for the representation of the protocol.   From the Langevin equation, we
have  the energy balance equation
\begin{equation}
\fl 
U(x(\tau); \nu_{\rm f})-U(x(-\tau); \nu_{\rm i})
= \int_{-\tau}^\tau dt \dot \nu \pder{U(x;\nu)}{\nu}+ 
\int_{-\tau}^\tau dt \dot x \circ \pder{U(x;\nu)}{x},
\end{equation}
where $\circ$ represents the Stratonovich rule of
the multiplication. The first term is interpreted as the work 
associated with the change in the parameter
and the second term corresponds to the heat
from the environment, which is denoted by $Q$. 
That is, for a given trajectory $\hat x \equiv (x(t))_{t=0}^\tau$,
we define the heat as 
\begin{equation}
Q (\hat x;\hat \nu) 
\equiv   \int_{-\tau}^\tau dt \dot x \circ \pder{U(x;\nu)}{x},
\label{heat}
\end{equation}
which was first pointed out in Ref. \cite{Sekimoto}.
Now, suppose that the system is in the equilibrium state at $t=0$. 
It is then shown that 
\begin{equation}
S(\nu_{\rm f})-S(\nu_{\rm i}) \ge  \beta \bra Q (\hat \nu) \ket,
\label{eq-2nd}
\end{equation}
where $\bra \ \ket $ represents the average of many realizations
for the given protocol $\hat \nu$.


We present a proof of (\ref{eq-2nd}) by using the transition
probability $\Psi$. We notice the identity of the type
\begin{equation}
\fl
\int dx_0 \int dx_1\cdots \int dx_{K} \
G(\hat x; \hat \alpha)  \prod_{k=0}^{K-1}
\Psi(x_k\to x_{k+1};\alpha_k) P_{\rm st}(x_0;\alpha_{\rm i})=1.
\label{identity}
\end{equation}
As one example of such $G$, we can choose 
\begin{equation}
G= 
\frac{P_{\rm st}(x_K;\alpha_{\rm f})}{P_{\rm st}(x_0;\alpha_{\rm i})}
\prod_{k=0}^{K-1}
\frac{\Psi(x_{k+1} \to x_{k};\alpha_k)}{\Psi (x_k \to x_{k+1};\alpha_k)}.
\label{FT-G}
\end{equation}
Indeed, by substituting (\ref{FT-G}) into the left-hand side of 
(\ref{identity}), we can confirm the identity (\ref{identity}), 
which is expressed as 
\begin{equation}
\bra e^{-\Sigma} \ket=1
\label{FT}
\end{equation}
in terms of the total entropy production  $\Sigma$ defined by
\begin{equation}
\Sigma \equiv  -\log P_{\rm st}(x(\tau);\alpha_{\rm f})+
          \log P_{\rm st}(x(-\tau);\alpha_{\rm i}) 
-\beta  Q(\hat x; \hat \alpha).
\end{equation}
From $e^{-x}\ge 1-x$, (\ref{FT}) leads to (\ref{eq-2nd}).
The identity (\ref{FT}) is equivalent to the Jarzynski 
equality and  it is also called an integral fluctuation
theorem. 


Here, from (\ref{can}), we have the adiabatic theorem
\begin{equation}
\pder{F}{\nu}=\bra \pder{U}{\nu} \ket
\label{adiabatic-eq}
\end{equation}
in equilibrium states. Since this can be rewritten as
\begin{equation}
\pder{S}{\nu}=\beta\left[\pder{\bra U \ket}{\nu} -\bra \pder{U}{\nu} 
\ket \right],
\label{adiabatic-eq-2}
\end{equation}
we can derive
\begin{equation}
S(\nu_{\rm f})-S(\nu_{\rm i}) =  \beta \bra Q (\hat \nu) \ket,
\label{eq-2nd-eq}
\end{equation}
for quasi-static operations $\hat \nu$ which are realized 
by a chain of infinitely small step operations. It should be
noted that the equality (\ref{eq-2nd-eq}) is related to the 
reversibility, because for the time-reversed protocol of $\hat \nu$,
which is defined by $\hat \nu^\dagger(t)=\nu(-t)$, (\ref{eq-2nd-eq}) 
leads to the reversibility 
\begin{equation}
\bra Q (\hat \nu^\dagger)\ket = -\bra Q (\hat \nu) \ket
\label{reverse}
\end{equation}
for quasi-static operations $\hat \nu$. Without the reversibility 
(\ref{reverse}), the equality part in (\ref{eq-2nd}) does not hold.


Although we have focused on the case that $\beta$ is constant 
in time, it is easy to extend  the argument so as to derive
the Clausius relation for the protocol in which  $\beta$ is 
changed as a function of time. 
 

Now, we consider the case $f \not =0$.  
In this case, the stationary distribution $P_{\rm st}(x;\alpha)$ is not
canonical. Nevertheless, suppose that the entropy $S$ is 
given by the Shannon entropy of $P_{\rm st}(x;\alpha)$: 
\begin{equation}
S(\alpha) \equiv - \int dx P_{\rm st}(x;\alpha) \log P_{\rm st}(x;\alpha).
\end{equation}
The heat, the energy transfer from the bath, is given by  
\begin{equation}
Q (\hat x;\hat \alpha) \equiv   \int_{-\tau}^\tau dt \dot x \circ
\left( \pder{U(x;\nu)}{x}-f \right).
\label{heat-ness}
\end{equation}
Since we can derive (\ref{FT}) even  for this case by using 
the same method, we obtain
\begin{equation}
S(\alpha_{\rm f})-S(\alpha_{\rm i}) \ge  \beta \bra Q (\hat \alpha) \ket
\label{eq-2nd-ness}
\end{equation}
for general cases. This implies the second law of 
thermodynamics for  non-equilibrium states. However, 
since $\bra Q (\hat \alpha^\dagger) \ket  \not =  
-\bra Q (\hat \alpha) \ket$ for 
general quasi-static operations $\hat \alpha$, 
the equality in (\ref{eq-2nd-ness}) does not hold for 
any quasi-static operations with $f \not =0$. 
Therefore,  the entropy  in non-equilibrium steady states 
with $f \not =0$ cannot be determined from the heat as Clausius did.


These considerations lead to the following explicit question 
for models that satisfy (\ref{LDB}):
(i) Find a renormalized heat $Q^{\rm ren}$  that satisfies 
\begin{equation}
\int d'Q^{\rm ren} =0 
\label{eq-2nd-sst-0}
\end{equation}
along  closed paths in the extended state space with $\beta$ fixed.
 (ii) Investigate whether or not the state variable ${\cal S}$ 
defined by $Q^{\rm ren}$  satisfies the inequality 
\begin{equation}
{\cal S}(\alpha_{\rm f})-{\cal S}(\alpha_{\rm i}) \ge  \beta 
\bra  Q^{\rm ren} (\hat \alpha) \ket
\label{eq-2nd-sst}
\end{equation}
for any processes given by $\hat \alpha$. (iii) Derive
a statistical expression of ${\cal S}(\alpha)$.


\section{Result}\label{result}


\subsection{dual system}


A key step  in solving the question is to find a new identity,
which may be similar to  (\ref{FT}), but may  be defined for
a quantity reversible in quasi-static operations. 
Recall that the irreversibility 
$\bra Q (\hat \alpha^\dagger) \ket  \not =  
-\bra Q (\hat \alpha) \ket$ for general quasi-static operations $\hat \alpha$
originates from the persistent heat generation (entropy production)
in non-equilibrium steady states.
We thus need to remove this persistent contribution. 
Our basic strategy, which was proposed in Ref. \cite{Hatano-Sasa},
is to employ the  decomposition of the force 
\begin{equation}
F(x;\alpha)=f-\pder{U(x;\nu)}{x}
\end{equation}
in the form
\begin{equation}
F(x;\alpha)=b(x;\alpha)-T \pder{\phi(x;\alpha)}{x}
\label{decom}
\end{equation}
using the individual entropy
\begin{equation}
\phi(x;\alpha)\equiv  -\log P_{\rm st}(x;\alpha).
\end{equation}
That is, we rewrite (\ref{Lan}) as 
\begin{equation}
\gamma \dot x= b(x;\alpha)-T\pder{\phi(x;\alpha)}{x}+\xi.
\label{Lan2}
\end{equation}
One may see that (\ref{decom}) is nothing but 
the definition of $b(x;\alpha)$,
but as described in Ref. \cite{Hatano-Sasa}, $b(x;\alpha)$ and 
$-T\partial_x \phi(x;\alpha)$ in (\ref{decom}) correspond 
to the irreversible  and reversible part of the force, 
respectively. We shall explain the origin of these names.


As a basic property of fluctuation, we review the concept
of ``duality''.  We consider the ensemble of trajectories 
$\hat x =(x(t))_{t=-\tau}^\tau$ in  non-equilibrium steady states 
with the parameter fixed.  
We denote the probability measure of trajectories as ${\cal P}(\hat x)$.
For this ensemble, we can consider the ensemble consisting of
time reversed trajectories $\hat x^\dagger$ such that $x^\dagger(t) =x(-t)$.
We then ask a stochastic process that yields the ensemble 
of $\hat x^\dagger$. Formally, such a stochastic process, which is 
called a dual process, generates the probability measure of 
trajectories ${\cal P}^\dagger(\hat x) = {\cal P}(\hat x^\dagger)$. 
We can prove  that for the Langevin equation (\ref{Lan}),
the dual process is given by the Langevin equation in the form
\begin{equation}
\gamma \dot x= -b(x;\alpha)-T\pder{\phi(x;\alpha)}{x}+\xi.
\label{Lan-dual}
\end{equation}
By comparing (\ref{Lan2}) and (\ref{Lan-dual}),
one finds that $b(x)$ and $-T\partial_x\phi(x)$ in the force decomposition 
(\ref{decom}) correspond to the irreversible and reversible part,
respectively.  


We present a proof. For the Langevin equation (\ref{Lan2}),
the probability density of trajectories with $x(-\tau)$ fixed 
is written as 
\begin{equation}
\Psi(\hat x)= C \e^{
-\frac{\beta}{4 \gamma} 
\int_{-\tau}^\tau dt 
[( \gamma \dot x-b+T \partial_x\phi)^2
-\frac{2}{\beta}\partial_x(T\partial_x\phi-b)]},
\label{Psi}
\end{equation}
where $C$ is the normalization constant. 
(See Appendix in Ref. \cite{Iwata-Sasa} for a derivation through the 
naive discretization of time.)
Let $\tilde \Psi$ be the probability of trajectories
for the Langevin equation (\ref{Lan-dual}). That is,
\begin{equation}
\tilde  \Psi(\hat x)= C \e^{
-\frac{\beta}{4 \gamma} 
\int_{-\tau}^\tau dt 
[( \gamma \dot x+b+T \partial_x \phi)^2
-\frac{2}{\beta}\partial_x(T\partial_x\phi+b)]}.
\end{equation}
By taking the ratio of $\Psi(\hat x)$ and $\tilde \Psi(\hat x^\dagger)$, 
we obtain
\begin{equation}
\frac{\Psi(\hat x)}{\tilde \Psi( \hat x^\dagger)}
=
\e^{-\int_{-\tau}^\tau dt 
[\dot x \partial_x \phi+(\partial_xb-b\partial_x\phi)/\gamma]}.
\end{equation}
From the steady state condition $b \e^{-\phi}=\gamma J={\rm const}$,
we have $\partial_xb=b\partial_x\phi$.  Thus, 
\begin{equation}
\Psi(\hat x) 
P_{\rm st}(x(-\tau)) 
=
\tilde  \Psi( \hat x^\dagger) P_{\rm st}(x(\tau)) .
\label{dual-result}
\end{equation}
This relation indicates that $\tilde  \Psi$ is
the dual transition probability of $\Psi$,
because 
${\cal P}^\dagger(\hat x)
= \tilde \Psi(\hat x) P_{\rm st}(x(-\tau))$.
We thus conclude that (\ref{Lan-dual}) is the Langevin equation
describing the dual process. As seen in the proof, 
the so-called Jacobian term ${2}\partial_x(T\partial_x\phi-b)/\beta$
in the path integral expression 
(\ref{Psi}) is inevitable to obtain (\ref{dual-result}).

\subsection{extended Clausius relation}


Now, we consider an external operation with a protocol $\hat \alpha$. 
For a given trajectory $\hat x$,  the heat (from the heat bath) is 
given by 
\begin{equation}
Q (\hat x;\hat \alpha) 
\equiv  \int_{-\tau}^\tau dt 
\dot x \circ  \left( T \pder{\phi(x;\alpha)}{x}-b \right).
\label{Q1}
\end{equation}
Suppose that this trajectory $\hat x$ is also observed in 
the dual Langevin equation. Then, the heat is 
\begin{equation}
Q^\dagger (\hat x;\hat \alpha) 
\equiv  \int_{-\tau}^\tau dt 
\dot x  \circ \left( T \pder{\phi(x;\alpha)}{x}+b \right).
\label{Q2}
\end{equation}
By expressing the Shannon entropy of $P_{\rm st}$ as 
\begin{equation}
S(\alpha) = \int dx \phi(x;\alpha)P_{\rm st}(x;\alpha),
\label{shannon-ness}
\end{equation}
we can prove 
\begin{equation}
S(\alpha_{\rm f})-S(\alpha_{\rm i}) 
\ge \beta (\bra Q \ket+ \bra Q^\dagger \ket)/2, 
\label{ex-clausius}
\end{equation}
where the equality holds for the quasi-static processes. 
This is the extended Clausius inequality which
is valid even  far from equilibrium. The result gives the 
answer to (\ref{eq-2nd-sst}) by setting $Q^{\rm ren}=(Q+Q^\dagger )/2$
and ${\cal S}(\alpha) =S(\alpha)$.


The proof of (\ref{ex-clausius}) is the following. We first define 
\begin{equation}
Y \equiv  \int dt \dot \alpha \pder{\phi(x;\alpha)}{\alpha}.
\end{equation}
From (\ref{Q1}) and (\ref{Q2}), we immediately obtain
\begin{equation}
\beta ( Q  + Q^\dagger )/2= \phi(x(\tau);\alpha_{\rm f})
-\phi(x(-\tau);\alpha_{\rm i})-Y.
\label{balance}
\end{equation}
This may be interpreted as a balance equation of 
the individual  entropy $\phi$. 
Next, by substituting 
\begin{equation}
G= 
\frac{P_{\rm st}(x_K;\alpha_{\rm f})}{P_{\rm st}(x_0;\alpha_{\rm i})}
\prod_{k=0}^{K-1}
\frac{P_{\rm st}(x_{k};\alpha_k)}{P_{\rm st}(x_{k+1};\alpha_k)},
\label{FT-HS}
\end{equation}
into the right-hand side of (\ref{identity}),
we can confirm the identity (\ref{identity}).
This is expressed as a non-equilibrium identity
\begin{equation}
\bra e^{-Y} \ket=1,
\label{HS}
\end{equation}
which  leads to a generalized second law 
\begin{equation}
\bra Y \ket \ge 0.
\label{g-2nd}
\end{equation} 
Since the adiabatic theorem in this case leads to $\bra Y \ket =0$
for quasi-static operations,  the equality in (\ref{g-2nd})
holds for quasi-static processes. 
By combining (\ref{balance}) with (\ref{g-2nd}),
we obtain (\ref{ex-clausius}). 


In the proof described above, one may doubt that 
(\ref{balance}) is also valid  for non-Langevin cases.
We thus present another proof so that we can consider general 
Markov processes. By using the steady state distribution 
$P_{\rm st}(x;\alpha_k)$, 
the dual transition probability $\Psi^\dagger(x_k \to x_{k+1};\alpha_k)$
is defined as
\begin{equation}
\Psi^\dagger (x_{k+1} \to x_{k};\alpha_k)
\equiv 
\frac{\Psi (x_{k} \to x_{k+1};\alpha_k) 
P_{\rm st}(x_{k};\alpha_{k})}{P_{\rm st}(x_{k+1};\alpha_{k})}.
\label{dual-def}
\end{equation}
From (\ref{LDB}) and (\ref{dual-def}), we obtain
\begin{equation}
\fl
\beta [Q(x_k \to x_{k+1};\alpha_k) +Q^\dagger (x_k \to x_{k+1};\alpha_k) ]/2
=  -  \log \frac{ P_{\rm st}(x_{k+1};\alpha_k)}{P_{\rm st}(x_k;\alpha_k) }.
\label{subst}
\end{equation}
The right-hand side is rewritten as 
\begin{equation}
\fl
-  \log P_{\rm st}(x_{k+1};\alpha_k) +  \log P_{\rm st}(x_{k+1};\alpha_{k+1}) 
-  \log P_{\rm st}(x_{k+1};\alpha_{k+1}) + \log P_{\rm st}(x_k;\alpha_k). 
\end{equation}
By taking the limit $K \to \infty$, we obtain (\ref{balance}).


We can rewrite (\ref{ex-clausius}) as another form.
Let us define $Q_{\rm hk}$ as 
\begin{equation}
Q_{\rm hk} \equiv (Q-Q^\dagger)/2,
\label{qhk}
\end{equation}
which is interpreted as an intrinsic part of 
irreversible heat. Then,  (\ref{ex-clausius})  becomes
\begin{equation}
S(\alpha_1)-S(\alpha_0) \ge \beta (\bra Q \ket - \bra Q_{\rm hk} \ket).
\label{ex-clausius-2}
\end{equation}
This is the expression proposed in Ref. \cite{Hatano-Sasa}. 
Motivated by Ref. \cite{Oono}, $Q_{\rm hk}$ was referred 
to as the house-keeping heat and  $Q-Q_{\rm hk}$ was called 
``excess heat''. However, this excess heat is different from
the more naive one given in  (\ref{excess}). 
More explicitly,  we can write 
$\bra  Q_{\rm hk}  \ket= -\int dt \bra \dot x(t) \circ b(x(t)) \ket$
whose integrand is different from the steady-state heat flux 
$-J_{\rm st} =\bra \dot x(t) \circ b(x(t)) \ket_{\rm st} $, where 
$\bra \ \ket_{\rm st} $ represents the expectation value in the stedy state
of the system with $\alpha(t)$ fixed virtually. It should be noted
that there is no  difference when we do not change the parameter
$\alpha$ in time. 
Since  we did not have any knowledge on the nature of of 
house keeping heat for time-dependent cases, we identified (\ref{qhk}) 
to be a mathematical expression of the house-keeping heat in the
phenomenological proposal.


The identity (\ref{HS}) takes the same form as (\ref{FT}).
Interestingly, when we consider the decomposition 
\begin{equation}
\Sigma=Y+Z,
\label{add}
\end{equation}
$Z$ is equal to $-Q_{\rm hk}=-(Q-Q^\dagger)/2$, and $Z$ also satisfies
\begin{equation}
\bra e^{-Z} \ket=1,
\label{SS}
\end{equation}
which was presented by Speck and Seifert \cite{Speck-Seifert}. 
See also Ref. \cite{Esposito}. The identity is obtained by
choosing 
\begin{equation}
G= 
\prod_{k=0}^{K-1}
\frac{\Psi(x_{k+1} \to x_{k};\alpha_k)}{\Psi (x_k \to x_{k+1};\alpha_k)}
\frac{P_{\rm st}(x_{k+1};\alpha_k)}{P_{\rm st}(x_{k};\alpha_k)}
\label{FT-SS}
\end{equation}
in  (\ref{identity}). 
The decomposition (\ref{add}) that satisfies relations 
 (\ref{FT}), (\ref{HS}), and (\ref{SS}) seems to be rather
surprising, but we do not understand a  physical principle
behind this fact.

\subsection{cases with odd parity variables}


The inequality (\ref{ex-clausius}) can be derived for a 
wide class of systems, as is understood from the 
derivation method. However, unfortunately, there is 
a restriction of the application. The inequality (\ref{ex-clausius}) 
does not hold for systems with odd-parity variables.
In order to demonstrate the difficulty, we consider 
the under-damped version of the Langevin equation 
(\ref{Lan}), which takes the form
\begin{equation}
m \ddot x+\gamma \dot x= f-\pder{U(x;\nu)}{x}+\xi,
\label{u-Lan}
\end{equation}
where $m$ is the mass. The dynamical variable in this model
is $z=(x,p)$ with $p=m \dot x$. Since $z$ becomes $z^*=(x,-p)$ 
for the time reversal operation,
$z$ contains the odd parity variable. For the trajectory 
$\hat z =(z(t))_{t=-\tau}^\tau$, the time reversed trajectory
$\hat z^\dagger$ is given by $z^\dagger(t)=z^*(-t)$. 
For this model, we have a transition probability 
$\Psi(z_k \to z_{k+1};\alpha_k) $
in the discrete time description. Then, the local detailed
balance condition (\ref{LDB}) is replaced by
\begin{equation}
-\beta Q(z_k \to z_{k+1};\alpha_k) 
= \log \frac{\Psi(z_k \to z_{k+1};\alpha_k)}{\Psi(z_{k+1}^* \to z_{k}^*;\alpha_k)}.
\label{LDB2}
\end{equation}
All the arguments in this subsection apply to general 
Markov processes satisfying (\ref{LDB2}).


Even in this case, the non-equilibrium equality (\ref{HS}) 
is valid, and this  leads to the generalized second law (\ref{g-2nd}). 
However, its 
energetic interpretation is not clearly obtained, as pointed
out by Refs. \cite{Spinny-Ford, Lee-Kwan-Park}. Following
our basic strategy  respecting the reversible part of heat,
we expect that $Q+Q^\dagger$ is related to the existence
of a state variable. Here, the
dual system is defined as that yielding the ensemble of
the reversed path $\hat z^\dagger$ starting from $z^{*}$,
where $z$ is taken from the stationary distribution 
of the system with $\alpha_{\rm f}$.  That is, the transition
probability of the dual system is defined as 
\begin{equation}
\Psi^\dagger (z_{k+1}^* \to z_{k}^*;\alpha_k)
\equiv 
\frac{\Psi (z_{k} \to z_{k+1};\alpha_k) 
P_{\rm st}(z_{k};\alpha_{k})}{P_{\rm st}(z_{k+1};\alpha_{k})}.
\label{dual-def2}
\end{equation}
Then, (\ref{subst}) is replaced by
\begin{equation}
\fl
\beta [Q(z_k \to z_{k+1};\alpha_k) +Q^\dagger (z_k \to z_{k+1};\alpha_k) ]
=  - \log 
\frac{ P_{\rm st}(z_{k+1};\alpha_k)P_{\rm st}(z_{k+1}^*;\alpha_k)}
{P_{\rm st}(z_k;\alpha_k) P_{\rm st}(z_k^*;\alpha_k)}.
\label{subst2}
\end{equation}
It should be noted that $P_{\rm st}(z) \not = P_{\rm st}(z^*)$
in this example and that this property holds for many systems. 
Obviously, (\ref{subst}) holds for some examples that 
satisfy $P_{\rm st}(z)  = P_{\rm st}(z^*)$. 
By setting 
\begin{equation}
\phi^{\rm sym}(z,\alpha)\equiv-\frac{1}{2} 
[\log P_{\rm st}(z;\alpha)+\log P_{\rm st}(z^*;\alpha)],
\end{equation}
we define
\begin{equation}
S^{\rm sym}(\alpha)
\equiv   \int dz  P_{\rm st}(z) \phi^{\rm sym}(z;\alpha)
\label{Shannon-ss-2}
\end{equation}
and
\begin{equation}
Y^{\rm sym}\equiv 
\int d\alpha \pder{\phi^{\rm sym}(z;\alpha)}{\alpha}.
\end{equation}
Now,  taking the limit $K \to \infty$ in (\ref{subst2}),
we obtain 
\begin{equation}
\beta (\bra Q \ket + \bra Q^\dagger \ket)/2
= S^{\rm sym}(\alpha_{\rm f})
-S^{\rm sym} (\alpha_{\rm i}) -\bra Y^{\rm sym} \ket.
\label{balance2}
\end{equation}
Since $Y^{\rm sym}$ is not related to $Y$, we cannot
combine the generalized second law (\ref{g-2nd}) 
with (\ref{balance2}). 


Nevertheless, when we focus on quasi-static processes near equilibrium,
we can rewrite (\ref{balance2}) as a stimulating form. Explicitly,
we consider a step process $\alpha(t)= \alpha_0+ \delta \alpha_1\theta(t)$,
where $\alpha_0$ and $\alpha_1$ are two parameter values, and 
$\theta(\ )$ is the Heaviside step function. We also set $\ep=\beta fL$.
We then assume that dimensionless quantities $\epsilon$ and $\delta$ 
are small. 
In the step process, we have
\begin{equation}
\bra Y^{\rm sym} \ket= \delta \alpha_1 \int dz P_{\rm st}(z;\alpha)
\pder{\phi^{\rm sym}(z;\alpha)}{\alpha}.
\end{equation}
By using the equality
\begin{equation}
\int dz P_{\rm st}(z;\alpha)\pder{\phi(z;\alpha)}{\alpha}=0,
\end{equation}
we obtain
\begin{equation}
\bra Y^{\rm sym} \ket= 
\frac{\delta \alpha_1}{4}
\int dz 
\left[ 
P_{\rm st}(z;\alpha)
\pder{\phi(z^*;\alpha)}{\alpha}+(z \leftrightarrow z^*)
\right].
\label{y-sym-rep}
\end{equation}
By noting  $P_{\rm st}(z^*;\alpha)-P_{\rm st}(z;\alpha)=O(\epsilon)$,
we find that 
\begin{eqnarray}
\phi(z^*; \alpha)&=&  - \log [
P_{\rm st}(z; \alpha)+P_{\rm st}(z^*; \alpha)-P_{\rm st}(z; \alpha)] 
\nonumber \\
& =& 
- \log P_{\rm st}(z; \alpha)
- \log 
\left[1+ \frac{P_{\rm st}(z^*; \alpha)-P_{\rm st}(z; \alpha)}
{P_{\rm st}(z; \alpha)} 
\right]
\nonumber  \\
& =& 
 \phi(z; \alpha)
- \frac{P_{\rm st}(z^*; \alpha)-P_{\rm st}(z; \alpha)}
  {P_{\rm st}(z; \alpha)} +O(\ep^2).
\end{eqnarray}
This gives 
\begin{eqnarray}
\int dz P_{\rm st}(z;\alpha)
\pder{\phi(z^*;\alpha)}{\alpha}
&=&
-\int dz P_{\rm st}(z;\alpha)
\pder{}{\alpha}
\frac{P_{\rm st}(z^*; \alpha)}{P_{\rm st}(z; \alpha)}
+O(\ep^2) \nonumber \\
&= & -\int dz P_{\rm st}(z^*;\alpha)
\pder{\phi(z;\alpha)}{\alpha}
+O(\ep^2) .
\label{eq5}
\end{eqnarray}
By substituting (\ref{eq5}) into (\ref{y-sym-rep}), we obtain
\begin{equation}
\bra Y^{\rm sym} \ket= O(\epsilon^2\delta).
\label{Ysim-est}
\end{equation}
Here, we rewrite $\bra \ \ket$ as $\bra \ \ket_{\hat\alpha}$
in order to explicitly express the protocol dependence.
We then denote the expectation value with respect to 
${\cal P}^\dagger(\ ;\hat \alpha)$ by $\bra \ \ket_{\hat\alpha}^\dagger$.
From the definitions, we have $\bra Q^\dagger \ket_{\hat \alpha}=-
\bra Q^\dagger \ket^{\dagger}_{\hat \alpha^\dagger}$. Noting 
$\bra Q^\dagger \ket^\dagger =\bra Q^\dagger \ket =O(\ep^2)$
in steady states, we see
$\bra Q^\dagger \ket^{\dagger}_{\hat \alpha^\dagger}
= \bra Q \ket_{\hat \alpha^\dagger}+O(\ep^2\delta)$. 
These estimations lead to
\begin{equation}
 \bra Q^\dagger \ket_{\hat \alpha}
= -\bra Q \ket_{\hat \alpha^\dagger}+O(\ep^2\delta).
\label{Q-est}
\end{equation}
By substituting (\ref{Ysim-est}) and (\ref{Q-est}) into (\ref{balance2}),
we obtain
\begin{equation}
S^{\rm sym}(\alpha_1)-S^{\rm sym} (\alpha_0) 
=\beta(\bra Q \ket_{\hat \alpha} - \bra Q \ket_{\hat \alpha^\dagger})/2
+O(\ep^2\delta).
\label{KNST}
\end{equation}
This extended Clausius relation has the advantage that the right-hand side 
can be obtained by a heat measurement in experiments without
knowing  details of a  system.  It should be noted that
\begin{equation}
( \bra Q \ket_{\hat \alpha} - \bra Q \ket_{\hat \alpha^\dagger})/2
=Q^{\rm ex}+O(\delta^2).
\end{equation}
Therefore, (\ref{KNST}) is rewritten as 
\begin{equation}
dS^{\rm sym}= \beta d'Q^{\rm ex} +O(\ep^2),
\label{ssym-2nd}
\end{equation}
which gives the solution to (\ref{eq-2nd-sst-0}) and the equality
part in (\ref{eq-2nd-sst})  by setting ${\cal S}=S^{\rm sym}$ and 
$Q^{\rm ren}=Q^{\rm ex}$ near equilibrium. However, the inequality in 
(\ref{eq-2nd-sst}) is not valid in this formulation. It should be
noted that (\ref{ssym-2nd}) holds for over-damped cases, where
$S^{\rm sym}$  becomes the Shannon entropy.


The relation (\ref{KNST}) with the symmetrized Shannon entropy
was proposed in Refs. \cite{KNST, KNST2} and developed further 
in Ref. \cite{Nakagawa}. (See also Ref. \cite{KNST3} for the 
mathematically rigorous derivation for Markov jump processes.)
The relation (\ref{KNST}) can be derived for (maybe) all 
non-equilibrium steady state system near equilibrium including 
heat conduction systems and sheared systems.

\section{Concluding remarks}


In this paper, we have discussed a possible framework of 
steady state thermodynamics on the basis of a review of
the Hatano-Sasa relation. In particular, a technically important
message  is  that the extended Clausius relation in the form 
 (\ref{ex-clausius}) can be understood from two identities, 
the  balance equation (\ref{balance}) and the generalized 
Jarzynski equality (\ref{HS}). When we consider  systems with
odd parity variables, (\ref{HS}) is still valid, but (\ref{balance})
is modified as (\ref{balance2}) which contains $Y_{\rm sym}$
that cannot be connected to (\ref{HS}). Nevertheless,
for quasi-static processes near equilibrium, only the condition 
(\ref{balance2}) gives the definition of the state variable in terms
of the excess heat that can be measured in a  calorimetric 
experiment. 


Recently, Bertini et al have proposed a different type of
extension of Clausius inequality  by studying fluctuating hydrodynamics
for a  density field. Their formulation utilizes the decomposition
similar to (\ref{decom}) and the work associated with the anti-symmetric 
current was subtracted. Although it shares common concepts 
with our formulation, there might be essential difference in the origin
of the inequality. For the moment, their inequality 
can be derived  only for a special type of fluctuating hydrodynamics. 
It might be interesting to uncover  a  universal structure behind 
their formulation. 


As a different recent approach, Maes and  Neto\v{c}n\'{y} 
have  formulated an  extended Clausius relation 
in  connecting with  dynamical fluctuation
theory \cite{Christian}. Their key concept is a modified system 
in which a time-dependent distribution becomes stationary. Since 
a framework using such a modified system often appear in 
recent non-equilibrium statistical mechanics 
\cite{Jack-Sollich, Nemoto-Sasa}, 
there might be one direction which we should consider seriously.


Furthermore, by considering an extension of the axiomatic formulation
of the thermodynamic entropy \cite{LY}, Lieb and Yngvason have argued 
non-equilibrium entropy from the viewpoint of adiabatic accessibility
\cite{LY2}. Here, it should be noted that
a set of axioms of ``adiabatic processes'' 
formulated in Ref. \cite{LY}  precisely specifies  real adiabatic processes, 
while a set of axioms of ``adiabatic processes'' formulated in  
non-equilibrium state spaces might allow us to make  different models 
of ``adiabatic processes''.


What is the most promising approach?  At  present,
we do not have an answer. Nevertheless, if we respect the operational 
determination of the entropy, the Hatano-Sasa formulation using 
$ Q^\dagger$, the result by Bertini et al,  and the approach by 
Maes-Neto\v{c}n\'{y}  have serious difficulties. The only possible way 
in the operational framework may be to employ $d'Q^{\rm ex}$, but we 
know that the thermodynamic framework 
on the basis of $d'Q^{\rm ex}$  may be restricted to a class of systems
near equilibrium. If we assume that $d'Q^{\rm ex}$ plays an essential role,
we should have another method. One  possible extension in this direction 
is to introduce the integration factor such that
\begin{equation}
d {\cal S} =  d'Q^{\rm ex}/T^{\rm eff}.
\end{equation}
The previous result 
indicates that $T^{\rm eff}=T$ and ${\cal S}=S^{\rm sym}$ near equilibrium. 
As far as we know,  $T^{\rm eff}$ and ${\cal S}$ have not been calculated
in this approach. It would be interesting if one could find 
the physical significance of  $T^{\rm eff}$.  


All the arguments in this paper are too formal. The most important 
question to be solved may be to find phenomena, of which an  extended 
form of the entropy provides a useful understanding. As one example,
Sasa and Tasaki studied  a force arising from the change in 
the statistical weight
in non-equilibrium steady states on the basis of 
a phenomenological framework of steady state thermodynamics 
\cite{Sasa-Tasaki}. Although this proposal was too naive, the direction
of thinking might be correct. It would be  amazing to find 
an experimental configuration that extracts purely statistical 
mechanical effects in non-equilibrium steady states. 

\ack

The author thanks T. Hatano, T. S. Komatsu, N. Nakagawa, and H. Tasaki
for their collaborations on the works reported in this paper.
He also thanks T. Nemoto, N. Shiraishi and Y. Oono 
for their useful comments on the manuscript. 
The present study was supported by KAKENHI Nos. 22340109, 25103002,
and by the JSPS Core-to-Core program 
``Non-equilibrium dynamics of soft-matter and information''.

\end{document}